\documentclass[aps,preprint]{revtex4}
\usepackage{amssymb}

\begin{document}

\author{Peng-ming Zhang$^1$\footnote{Author to whom
correspondence should be addressed. Email: zhpm@lzu.edu.cn},
Li-ming Cao$^2$, Yi-shi Duan$^2$ and Cheng-kui Zhong$^1$}
\title{Transverse force on a moving vortex with the acoustic geometry}
\date{\today}

\begin{abstract}
We consider the transverse force on a moving vortex with the acoustic metric
using the $\phi $-mapping topological current theory. In the frame of
effective spacetime geometry the vortex appear naturally by virtue of the
vortex tensor in the Lorentz spacetime and we show that it is just the
vortex derived with the order parameter in the condensed matter. With the
usual Lagrangian we obtain the equation of motion for the vortex. At last,
we show that the transverse force on the moving vortex in our equation is
just the usual Magnus force in a simple model.
\end{abstract}

\affiliation{$^1$Department of Mathematics, Lanzhou University,
Lanzhou, 730000,
P. R. China\\
$^2$Institute of Theoretical Physics, Lanzhou University, Lanzhou,
730000, P. R. China}

\maketitle

\section{Introduction}

Recently the analog model is very popular for studying the
condensed matter physics (see \cite{Rev} for a review). Since
direct experimental probes of many important aspects of general
relativity (GR) are extremely difficult, the possibility of using
condensed matter system, such as Bose-Einstein condensates (BEC),
to mimic certain aspects of GR could prove to be very important
\cite{Garay,Garay2,Barcelo}. These analog models provide a bridge
for interchanging conceptualizations of phenomena between various
condensed matter systems and relativistic physics \cite
{Rev,Visser,VF,Unruh,Volovik,Garay,Garay2,Barcelo,Stone},
sometimes they illuminate aspects of general relativity and
sometimes the machinery of differential geometry can be used to
illuminate aspects of the analog model. In this letter, we use
mathematical methods developed in the framework of differential
geometry to study the transverse force on a moving vortex in
Bose-Einstein condensates. On the other hand, we have to mention
that these condensed matter systems can also be used to simulate
topological defects characteristic of gauge theories and which are
considered to have played a cosmological role in the early stages
of the evolution of the universe such as monopoles and cosmic
strings.

In the analog model, what we are concerned is the propagation of
small collective perturbations of the condensate around a
background stationary state instead of solving the
Gross-Pitaevskii (GP) equation with some given external potential.
By virtue of the idea that an effective Lorentzian metric governs
perturbative fluctuations, analog models can be given based on
acoustic propagation in an irrotational vortex. Thus we can use
the method of general relativity to investigate the ''effective
space-time geometry'' in a constant-speed-of-sound (iso-tachic)
and almost incompressible\ (iso-pycnal) hydrodynamical flows
\cite{VF}. With the so-called ''effective acoustic metric'' the
condensates can be regards as Lorentz spacetime \cite
{Garay,Garay2,Barcelo}. It is natural that the cosmic string in
the Lorentz spacetime will appear in such effective spacetime
geometry as the vortex. Furthermore, in terms of the
energy-momentum tensor we consider the equation of motion for the
vortex and calculate the transverse force on a moving vortex in
detail. We conclude that the Magnus force can be described with
the effective acoustic metric in the frame of general relativity
without any concrete model or hypothesis.

\section{Topological vortex in the BEC}

Given the effective acoustic metric, the effective gravity arises
in the BEC system \cite{Garay,Garay2,Barcelo}, i.e. we can
consider such system as the Lorentz spacetime as follows.

Bose--Einstein condensates are most usefully described by the nonlinear
Schr\"{o}\-dinger equation, also called the Gross--Pitaevskii equation:
\begin{equation}
i\hbar \frac \partial {\partial t}\psi (t,\vec{x})=(-\frac{\hbar ^2}{2m}%
\nabla ^2+V_{ext}(\vec{x})+\lambda |\psi (t,\vec{x})|^2)\psi (t,\vec{x}).
\label{lg1}
\end{equation}
Now use the Madelung representation~\cite{Madelung} to put the
Schr\"{o}dinger equation in ``hydrodynamic'' form:
\begin{equation}
\psi =\sqrt{\rho }\;\exp (-i\theta \;m/\hbar ).
\end{equation}
Take real and imaginary parts: The imaginary part is a continuity equation
for an irrotational fluid flow of velocity $\vec{v}\equiv \nabla \theta $
and density $\rho $; while the real part is a Hamilton--Jacobi equation
(Bernoulli equation; its gradient leads to the Euler equation).
Specifically:
\begin{equation}
\partial _t\rho +\nabla \cdot (\rho \;\nabla \theta )=0.
\end{equation}
\begin{equation}
\frac \partial {\partial t}\theta +{\frac 12}(\nabla \theta )^2+{\frac{%
\lambda \;\rho }m}-{\frac{\hbar ^2}{2m^2}}\;{\frac{\Delta \sqrt{\rho }}{%
\sqrt{\rho }}}=0.
\end{equation}
That is, the nonlinear Schr\"{o}dinger equation is completely equivalent to
irrotational inviscid hydrodynamics with a particular form for the enthalpy
\begin{equation}
h=\int \frac{dp}\rho =\frac{\lambda \rho }m,
\end{equation}
plus a peculiar derivative self-interaction:
\begin{equation}
V_Q=-{\frac{\hbar ^2}{2m^2}}\;{\frac{\Delta \sqrt{\rho }}{\sqrt{\rho }}}.
\end{equation}
The equation of state for this ``quantum fluid'' is calculated from the
enthalpy
\begin{equation}
p={\frac{\lambda \;\rho ^2}{2m}}.
\end{equation}
The corresponding speed of sound is
\begin{equation}
c_s^2=\frac{dp}{d\rho }=\frac{\lambda \rho }m.
\end{equation}

The disturbances propagate in an effective spacetime with metric $g_{\mu \nu
}$, which was shown to be of the Painleve-Gullstrand form \cite{Visser,VF}
\[
g_{00}=-\frac \rho c[c^2-v^2],\;\;g_{0i}=-\frac \rho cv_i,\;\;g_{ij}=\frac
\rho c\delta _{ij},
\]
where the velocity $c$ plays the role of the speed of light and is equal to
the sound speed for phonons. The metric has spacetime interval
\[
ds^2=\frac \rho c[-c^2dt^2+\delta _{ij}(dx^i-v^idt)(dx^j-v^jdt)],
\]
where the indices on the background velocity $v^i$ are always raised and
lowered using the flat 3-dimensional Cartesian metric, i.e. $v^i=v_i$ and $%
v^2=v^iv_i$.

Consider the physical situation that the speed of sound is iso-tachic and
independent of position and time, we can choose co-ordinates to set the
speed $c$ of linear quasiparticle dispersion equal to unity. The
(3+1)-dimensional Painleve-Gullstrand metric then reads \cite{VF}
\[
g_{\mu \nu }=\rho \left[
\begin{array}{ll}
-1+v^2 & \;-\vec{v} \\
\;-\vec{v} & \;\;\;1
\end{array}
\right]
\]
The inverse metric is
\[
g^{\mu \nu }=\frac 1\rho \left[
\begin{array}{ll}
-1 & \;\;\;\;-v^j \\
-v^i & \;\;\delta ^{\iota j}-v^iv^j
\end{array}
\right]
\]

Since the general relativity can be described with the tetrad field as the
SO(3,1) gauge theory, the above BEC system should possess the similar
effective tetrad field $e_\mu ^a$ ($a$ and $\mu$ are SO(3,1) and space-time
indices, respectively). As we known, $\omega _{\mu ab}$ is the connection of
Lorentz group gauge theory
\[
D_\mu \phi _a=\partial _\mu \phi _a-\omega _{\mu ab}\phi _b,\;\;\;\;\;\mu
=0,1,2,3,\;\;\;\;\;a,b=1,2,3,4
\]
and the corresponding pure connection is defined as
\[
\omega _{abc}=e_a^\mu \omega _{\mu bc}.
\]
With this connection the vortex tensor is proposed as
\[
F_{\mu \nu }=e_{\nu a}D_\mu \omega _a-e_{\mu a}D_\nu \omega _a
\]
where $\omega _a=\omega _{bab}$. For Lorentz spacetime, the torsion should
be zero, i.e.
\[
T_{\mu \nu a}=D_\mu e_{\nu a}-D_\nu e_{\mu a}=0,
\]
then, the vortex tensor
\begin{eqnarray}
F_{\mu \nu } &=&\partial _\mu A_\nu -\partial _\nu A_\mu ,  \label{tensor1}
\end{eqnarray}
where $A_\mu =e_{\mu a}\omega _a$ can be looked as a U(1) connection.

In terms of Eqs. (\ref{tensor1}) the topological charge of vortex can be
found
\begin{equation}
q=\int_\Sigma F_{\mu \nu }dx^\mu \wedge dx^\nu .
\end{equation}
As one has shown in \cite{DuanU1}, the U(1) gauge potential can be
decomposed by the 2-dimensional unit vector fields $n^A$ $(n^A=\phi ^A/\sqrt{%
\phi ^B\phi ^B})$ as
\begin{equation}
A_\mu =\frac 1{2\pi }\varepsilon _{AB}n^A\partial _\mu n^B.
\end{equation}
One can find that the charge of vortex can be expressed as
\begin{equation}
q=\frac 1{2\pi }\int_\Sigma \varepsilon _{AB}\partial _\mu n^A\partial _\nu
n^Bdx^\mu \wedge dx^\nu .
\end{equation}
Following the $\phi $-mapping theory, it can be rigorously proved that
\begin{equation}
q=\int_\Sigma \delta ^2(\vec{\phi})D_{\mu \nu }(\frac \phi x)dx^\mu \wedge
dx^\nu ,
\end{equation}
where $\Sigma $ is an arbitrary 2-dimensional surface,
\begin{equation}
x^\mu =x^\mu (u^1,u^2),
\end{equation}
and $D^{\mu \nu }(\frac \phi x)$ is the tensor Jacobian which defined as
\begin{equation}
\varepsilon ^{AB}D^{\mu \nu }\left( \frac \phi x\right) =\varepsilon ^{\mu
\nu \lambda \sigma }\partial _\lambda \phi ^A\partial _\sigma \phi ^B.
\end{equation}
The integral $q$ can be rewritten with the usual Jacobian $D(\phi /u)$
\begin{equation}
q=\int_\Sigma \delta ^2(\vec{\phi})D(\frac \phi u)\sqrt{g_u}d^2u.
\end{equation}
We find that $q\neq 0$, only when
\begin{equation}
\phi ^A(\vec{x},t)=0,\;\;\;\;\;\;A=1,2.
\end{equation}
Then solutions of above equations are vortices
\begin{equation}
S_\alpha :\;\;\;\;\;\;x^\mu =z_A^\mu (\sigma ,\tau ),\;\;\;\;\;\;\alpha
=1,2,...,l
\end{equation}
which are the worldlines of vortices.

Using the $\phi $-mapping topological current theory we have
\begin{equation}
q=\int_\Sigma \sum_{A=1}^lW_\alpha \delta ^2(u^i-z_\alpha ^i)d^2u
\end{equation}
where $z_\alpha ^i(\alpha =1,2,...,l)$ are the intersection points of
vortices $S_\alpha $ with surface $\Sigma $ and $W_\alpha $ is the winding
number. Then we find
\begin{equation}
q=\sum_{\alpha =1}^lW_\alpha .
\end{equation}
This is just our $\phi $-mapping topological current theory of
vortices in the frame of effective acoustic geometry which shows
that the vortices appear naturally in the hydrodynamical flows and
the charge of vortices are topologically quantized by winding
number. In the other word, the vortices will emerge in all those
systems which can be described with the effective Lorentz geometry
with the above discussion.

At the other hand, the vortex can be discussed with the order parameter $%
\psi (\vec{r},t)$ directly. Starting from the Gross-Pitaevskii equation, one
can construct the vortex current with the condensed wave function $\psi $
\[
\vec{j}=\frac m\hbar \nabla \times \vec{v},
\]
where the current velocity
\[
\vec{v}=-\frac{i\hbar }{2m}(\psi ^{*}\nabla \psi -\psi \nabla \psi
^{*})/|\psi |^2,
\]
is just the background velocity.

It is well known that the condensed wave function $\psi $ can be looked upon
as a section of a complex line bundle with base manifold $M$ (in this paper $%
M=R^3\otimes R$). Denoting the condensed wave function $\psi $ as
\[
\psi (\vec{x},t)=\phi ^1(\vec{x},t)+i\phi ^2(\vec{x},t),
\]
where $\phi ^1(\vec{x})$ and $\phi ^2(\vec{x})$ are two components of a
two-dimensional vector field
\[
\vec{\phi}=(\phi ^1,\phi ^2)
\]
in the (3+1)-dimensional space-time. The vortex current in the 3-dimensional
space can be obtained
\[
j^i=\frac 1{2\pi }\varepsilon ^{ijk}\varepsilon _{AB}\partial _\nu
n^A\partial _\lambda n^B,\;\;\;\;i,j,k=1,2,3
\]
where $n^A$ is the two-dimensional unit vector field of the complex scalar
field:
\[
n^A=\phi ^{\newline
A}/||\phi ||,\;\;\;\;||\phi ||^2=\phi ^A\phi ^A,\;\;\;A=1,2.
\]
It is clear that the topological current is identically conserved \cite
{Gelfand}, i.e.
\begin{equation}
\partial _ij^i=0.  \label{conserv1}
\end{equation}
By making use of the $\phi $-mapping method, this topological current can be
rewritten in a compact form \cite{DuanZhangLi},
\begin{equation}
j^i=D^i(\frac \phi x)\delta (\vec{\phi}),  \label{zero12}
\end{equation}
where $D^i(\frac \phi x)$ is the vector Jacobians of $\phi (x):$%
\[
D^i(\frac \phi x)=\frac 12\varepsilon ^{ijk}\varepsilon _{AB}\partial _j\phi
^A\partial _k\phi ^B.
\]
Thus we have the important relation between the topological current and the
condensed wave function $\psi (\vec{x})$ in the Bose-Einstein condensation
system. With this topological current the corresponding vorticity $\Gamma
=\oint \vec{v}\cdot d\vec{l}$ can be given
\[
\Gamma =\frac hm\sum_{\alpha =1}^lW_\alpha =q\frac hm.
\]
One can find easily that the vortex in the frame of the effective acoustic
geometry is the same vortex given with the order parameters. That means that
we can discuss this kinds of topological defects in the condensed matter
using the method of general relativity by virtue of the effective acoustic
metric.

It is convenient to generalize the vortex current into the (3+1)-dimensional
spacetime
\[
j^{\mu \nu }=\varepsilon ^{\mu \nu \lambda \rho }F_{\lambda \rho }.
\]
It is easy to prove that
\[
j^{\mu \nu }=\delta (\vec{\phi})D^{\mu \nu }(\phi /x),
\]
where $D^{\mu \nu }(\phi /x)$ is the tensor Jocobian
\[
\varepsilon _{\mu \nu \lambda \rho }D^{\mu \nu }(\phi /x)=\varepsilon
_{ab}\partial _\lambda \phi ^a\partial _\rho \phi ^b.
\]
Based on this generalized vortex current we can consider the
equation of emotion of vortex in the effective spacetime.

\section{Equation of emotion with the Energy-Momentum tensor}

In the above section, we show the vortices exist in the effective
acoustic spacetime geometry and their charges are quantized in the
level of topology. With the vortex current $j^{\mu \nu }$, we can
define the Lagrangian in the effective spacetime
\[
L=T\sqrt{\frac 12g_{\mu \nu }g_{\lambda \rho }j^{\mu \lambda }j^{\nu \rho }}%
=T\sqrt{\frac 12j_{\mu \nu }j^{\mu \nu }}
\]
which is the generalization of Nielsen's Lagrangian, where $T$ is
a constant with dimension of $[mass]^2$. It is easy to obtain the
energy-momentum tensor
\begin{equation}
T^{\mu \nu }=T\delta (\vec{\phi})D(\phi /x)g^{IJ}\frac{\partial x^\mu }{%
\partial u^I}\frac{\partial x^\nu }{\partial u^J},  \label{net1}
\end{equation}
which shows that the tensor only appear in the zeroes of the order parameter
field $\vec{\phi}(x)$, i.e. the position of the vortices. Since the vortex
is the only quasiparticles in the flowing fluids and offer the role of the
matter term in the effective acoustic geometry, Eq. (\ref{net1}) is natural
result for the energy-momentum tensor. From the principle of the least
action or the formula $\triangledown _\mu T^{\mu \nu }=0$ we can obtain the
equation of motion
\[
\frac 1{\sqrt{-g_u}}\frac \partial {\partial u^I}(\sqrt{-g_u}g^{IJ}\frac{%
\partial x^\lambda }{\partial u^J})+\Gamma _{\mu \nu }^\lambda g^{IJ}\frac{%
\partial x^\mu }{\partial u^I}\frac{\partial x^\nu }{\partial u^J}=0,
\]
which is the basic equations for us to discuss the transverse force on the
moving vortex. If we choose the conformal gauge, the equation have the
simple form
\begin{equation}
\partial _I\partial _Ix^\lambda +\Gamma _{\mu \nu }^\lambda \frac{\partial
x^\mu }{\partial u^I}\frac{\partial x^\nu }{\partial u^J}=0.  \label{eom1}
\end{equation}

In the following we consider a simple model that the system only include one
vortex and the background velocity is along with the $x$-direction, i.e. the
velocity $\vec{v}=(v_1,0,0)$. With this 3-velocity the connection
coefficients read
\begin{equation}
\Gamma _{00}^2=-\frac 12\partial _2v_1^2,\;\;\;\;\Gamma _{10}^2=\frac
12\partial _2v_1.  \label{connection2}
\end{equation}
Since we discuss the transverse force on the vortex, we choose an
ideal model that the vortex coordinate $x^\mu $ satisfies
\[
x^3=\sigma ,\;x^1=x^1(\tau ),\;x^2=x^2(\tau ),
\]
and the vortex moves in the $x$-direction only
\[
v_{vortex}=\dot{x}^1.
\]
Then the Eq. (\ref{eom1}) gives
\[
\partial _\tau \partial _\tau x^2+(\Gamma _{00}^2\partial _\tau x^0\partial
_\tau x^0+2\Gamma _{10}^2\partial _\tau x^1\partial _\tau x^0)=0,
\]
which can be calculate by virtue of the connection (\ref{connection2})
\[
\stackrel{..}{x}^2-v_1\partial _2v_1+\partial _2v_1\dot{x}^1=0,
\]
i.e.
\[
\stackrel{..}{x}^2=\partial _2v_1(v_1-v_{vortex}).
\]
Thus the transverse force $F_t$ on the vortex is obtained with the effective
acoustic geometry
\[
F_t=\rho _s\stackrel{..}{x}^2=\rho _s\Omega (v_{vortex}-v_n),
\]
where $\rho _s$ is the density of fluids, $\Omega =\partial _1v_2-\partial
_2v_1=-\partial _2v_1$ is the vorticity of the fluid and $v_n=v_1$ means the
background velocity of fluid. The last equation show that the vortex moving
in the fluids will be exerted a transverse force, just the Magnus force.

\section{Conclusion}

In this paper, we discuss the vortex in the BEC using the method
of effective acoustic geometry. Instead of solving the concrete GP
equation with some given external potential $ V_{ext}(x)$, we
study the topological structure of vortices from the viewpoint of
spacetime defects. We show that the vortex appears naturally in
such a effective Lorentz spacetime and the charge of vortex is
quantized in the topological level. Then, the energy-momentum
tensor is given with the vortex current in the effective
spacetime. The equation of vortex motion is derived in terms of
the energy-momentum tensor in the frame of general relativity.
Furthermore, we consider the transverse force on the vortex with
this equation of emotion. A simple model is calculated with the
expression of transverse force by virtue of the effective acoustic
metric. We find the transverse force in the frame of effective
geometry is just the usual Magnus force upon a moving vortex.

\section{Acknowledgement}

This project was supported in part by the National Natural Science
Foundation of China (NSFC-10175028), the TianYuan Mathematics Fund
(A0324661) and the China Postdoctoral Science Foundation.


\begin{thebibliography}{99}
\bibitem{Rev}  Workshop on ''Analog Models of General Relativity'' (Rio de
Janeiro, October, 2000); http://www.physics.wustl.edu/\symbol{126}%
visser/Analog;


\bibitem{Garay}  L. J. Garay, J. R. Anglin, J. I. Cirac and P. Zoller,
\textit{Black holes in Bose-Einstein condensates}, Phys. Rev. Lett. \textbf{%
85} (2000) 4643; [gr-qc/0002015];

\bibitem{Garay2}  L. J. Garay, J. R. Anglin, J. I. Cirac and P. Zoller,
\textit{Sonic black holes in dilute Bose-Einstein condensates},
Phys. Rev. A \textbf{63} (2001) 023611; [gr-qc/0005131];


\bibitem{Barcelo}  Carlos Barcelo, Stefano Liberati, and Matt Visser,
\textit{Analog gravity from Bose-Einstein condensates}, Class.
Quant. Grav. \textbf{18} (2001) 1137; gr-qc/0011026;


\bibitem{VF}  U. R. Fischer and M. Visser, \textit{Riemannian geometry of
irrotational vortex acoustics}, Phys. Rev. Lett. \textbf{88} (2002) 110201;
cond- mat/ 0110211; \textit{On the space-time curvature experienced by
quasiparticle excitations in the Painleve-Gullstrand effective geometry},
Ann. Phys. \textbf{304} (2003) 22-39; cond-mat/0205139;

\bibitem{Visser}  M. Visser, \textit{Acoustic black holes: Horizons,
ergospheres, and Hawking radiation} Class. Quantum Grav. \textbf{15} (1998)
1767-1791; gr-qc/9712010; \textit{Hawking radiation without black hole
entropy}, Phys. Rev. Lett. \textbf{80} (1998) 3436;

\bibitem{Unruh}  W. G. Unruh, \textit{Experimental black hole evaporation?}
Phys. Rev. Lett, \textbf{46} (1981) 1351--1353;

\bibitem{Volovik}  G. E. Volovik, \textit{Simulation of Painleve-Gullstand
Black Hole in thin }$^3$\textit{He-A film}, JETP Lett. \textbf{69} (1999)
705;

\bibitem{Stone}  M. Stone, \textit{Acoustic energy and momentum in a moving
medium}, Phys. Rev. B \textbf{62} (2000) 1341; \textit{Iordanskii force and
the gravitational Aharonov-Bohm effect for a moving vortex}, Phys. Rev. B
\textbf{61} (2000) 11780; cond-mat/9909313;

\bibitem{Madelung}  E. Madelung, \textit{Quantentheorie in hydrodynamischer
Form}, Zeitschrift f\"{u}r Physik \textbf{38}, 322 (1926).

\bibitem{DuanU1}  Y. S. Duan, in Proceedings of the Symposium on Yang-Mills
Gauge Theories, Beijing, 1984; Y. S. Duan, G. H. Yang and Y. Jiang, \textit{%
The origin and bifurcation of the space-time defects in the early Universe},
Gen. Rel. Grav. \textbf{29} (1997) 715.

\bibitem{Gelfand}  I. M. Gel'fand and G. E. Silov, \emph{Generalized Function%
}, Vol. 1 (Academic Press, New York, 1964).

\bibitem{DuanZhangLi}  Y.S. Duan, H. Zhang and S. Li, \textit{Topological
structure of the London equation}, Phys. Rev. B \textbf{58} (1998) 125.
\end{thebibliography}
\end{document}